\documentclass[final,5p,times,twocolumn]{elsarticle}
\usepackage{lineno,hyperref,amsmath,amsthm,amssymb,amsfonts,ragged2e,color,subfig}
\journal{``Physics"}
\begin{document}
\begin{frontmatter}
\title{Dust-acoustic envelope solitons and rogue waves in an electron depleted plasma}
\author{J. Akter$^{*,1}$, N. A. Chowdhury$^{**,1}$, and A. A. Mamun$^{1,2}$}
\address{$^{1}$Department of Physics, Jahangirnagar University, Savar, Dhaka-1342, Bangladesh\\
$^{2}$Wazed miah science research centre, Jahangirnagar University, Savar, Dhaka-1342, Bangladesh\\
Email: $^*$akter277phy@gmail.com, $^{**}$nurealam1743phy@gmail.com}
\begin{abstract}
Theoretical investigation of the nonlinear propagation and modulational
instability (MI) of the dust acoustic (DA) waves (DAWs) in an unmagnetized
electron depleted dusty plasma (containing opposite polarity warm dust grains and non-extensive positive ions)
has been made by deriving a nonlinear Schr\"{o}dinger equation with the help of perturbation method.
Two types of mode, namely, fast and slow DA modes, have been found.
The criteria for the formation of bright and dark envelope solitons as well as the first-order and second-order
rogue waves have been observed. The effects of various dusty plasma parameters (viz., dust mass, dust charge,
dust and ion number densities, etc.) on the MI of DAWs have been identified.
It is found that these dusty plasma parameters significantly modify the basic features of the DAWs.
The applications of the results obtained from this theoretical investigation in different regions
of space, viz., magnetosphere of Jupiter, upper mesosphere, Saturn's F-ring, and cometary tail, etc.
\end{abstract}
\begin{keyword}
Electron depletion \sep NLSE \sep modulational instability \sep envelope solitons \sep rogue waves.
\end{keyword}
\end{frontmatter}
\section{Introduction}
\label{1sec:Introduction}
The existence of highly charged massive dust grains does not only occurs in astrophysical
environments, viz., supernova explosion \cite{Sahu2012a}, Martian atmosphere \cite{Hossen2016b},
cometary tail \cite{Zaghbeer2014,Hossen2016a,Hossen2016b}, circumstellar clouds \cite{Hossen2017},
interstellar clouds \cite{Hossen2017}, solar system \cite{Hossen2017}, upper mesosphere \cite{Jahan2019,Zaghbeer2014},
magnetosphere of Jupiter \cite{Jahan2019}, and Saturn's F-ring \cite{Bains2013,Sahu2012b} but also in
many laboratory experiments, viz., laser-matter interaction \cite{Shahmansouri2013},
Q-machine \cite{Shukla2002}, ac-discharge \cite{Shukla2002}, rf-discharges \cite{Shukla2002}, fusion
devices \cite{Shukla2002}, etc. The investigation of nonlinear electrostatic pulses, viz., dust-acoustic (DA) double
layers (DADLs) \cite{Hossen2016a,Sahu2012b}, DA shock waves (DASHWs) \cite{Ferdousi2015,Hossen2017},
DA solitary waves (DASWs) \cite{Hossen2016a,Hossen2016b}, DA rogue waves (DARWs), and DA Gardner
solitons (DAGSs) \cite{Emamuddin2013} associated with DA Waves (DAWs) \cite{Barkan1995}  in multi-component plasma medium (MCPM)
within theoretical framework \cite{Ferdousi2017,Eghbali2017,Rao1990}
is one of the most interesting topics among the plasma physicists in twentieth century.

Non-extensive particles can be found in non-equilibrium complex systems due
to the existence of external force fields (viz., long range interactions, gravitational
forces, and Coulomb electric forces, etc.), and are also governed by the
non-extensive $q$-distribution which first introduced by Renyi \cite{Renyi1955}
and subsequently improved by Tsallis \cite{Tsallis1988}, and have been
identified by the GEOTAIL \cite{Eastman1998,Pavlos2011}. It may be noted that the
non-extensive parameter ($q$) describes the characteristics of the non-extensive $q$-distribution \cite{Rahman2018}.
A number of authors have considered non-extensive electrons \cite{Zaghbeer2014,Jahan2019,Bains2013,Eghbali2017,Rahman2018,Bacha2017} for interpreting the
characteristics of the nonlinear electrostatic waves in space environments,
viz., cometary tail \cite{Zaghbeer2014},  upper mesosphere \cite{Zaghbeer2014},
magnetosphere of Jupiter \cite{Jahan2019}, Saturn's F-ring \cite{Bains2013,Rahman2018},
ionosphere \cite{Eghbali2017}, lower part of magnetosphere \cite{Eghbali2017}, and solar wind \cite{Bacha2017}, etc.
Emamuddin \textit{et al.} \cite{Emamuddin2013} considered MCPM, and studied DAGSs in
presence of non-extensive plasma particles, and found that DAGSs exhibit positive and
negative solitons according to the critical value of $q$.
Roy \textit{et al.} \cite{Roy2014} examined the DASHWs in non-extensive plasmas, and
obtained that the positive DASHWs height decreases as $q$. Bacha and Tribeche \cite{Bacha2012}
demonstrated the amplitude of the pulse increases while the width of the pulse decreases
with increasing $q$ in a three components dusty plasma.

In many space and laboratory dusty plasma situations, most of the
background electrons could stick onto the surface of dust grains during the
charging processes and as a result one might encounter a significant depletion
of the electron number density in the ambient dusty plasma \cite{Sahu2012b,El-Tantawy2012,Borhanian2013}.
This scenario is relevant to a number of space dusty plasma systems, for example, planetary
rings (particularly, Saturn's F-ring), and laboratory experiments. It should be noted here that a complete depletion of
the electrons is not possible because the minimum value of the ratio between the electron and ion number densities
turns out to be the square root of the electron to ion mass ratio when electron and ion temperatures are approximately
equal and the grain surface potential approaches zero \cite{Sahu2012b,El-Tantawy2012,Borhanian2013}.
A number of authors have studied the propagation of the nonlinear electrostatic waves in electron depleted dusty plasma (EDDP) \cite{El-Tantawy2012,Borhanian2013,Tribeche2000,Tribeche2004,Tadsen2015,Steckiewicz2015,Shukla1992,Tagare1997}.
Borhanian and Shahmansouri \cite{Borhanian2013} studied three dimensional DASWs in an EDDP with two temperature
super-thermal ions, and highlighted that the super-thermality of ions leads to increase the amplitude of the solitary waves.
Tantawy and Moslem \cite{El-Tantawy2012} examined DASWs and DASHWs in a MCPM in presence of positive and negative ions,
and observed that the number density of the positive and negative ions as well as dust grains are significantly modified the amplitude
of DASWs and DASHWs. Shukla and Silin \cite{Shukla1992} investigated the features of the low-frequency dust-ion-acoustic
waves in an unmagnetized EDDP medium. Tagare \cite{Tagare1997} examined DASWs and DADLs in three components EDDP medium having cold dust grains and two temperature Maxwellian ions, and observed that DASWs exist in a particular region where DADLs do not exist.
Ferdousi \textit{et al.} \cite{Ferdousi2017} analyzed DASHWs in multi-component EDDP medium having non-thermal plasma species.
Sahu and Tribeche \textit{et al.} \cite{Sahu2012b} reported small amplitude DADLs in a two components
EDDP with ions featuring non-extensive distribution, and observed that the non-extensivity
of the ions admits compressive as well as rarefactive DADLs in EDDPs. Hossen \textit{et al.} \cite{Hossen2016a}
considered three components DP medium containing mobile opposite polarity dust grains (OPDGs) and non-extensive ions
to study DASWs, and found that the height of the positive solitary potential
increases with positive ion number density while the negative solitary potential decreases.

The nonlinear Schr\"{o}dinger equation (NLSE) is arguably the most far-reaching and
beautiful physical equation ever constructed \cite{C1,C2,C3,Kourakis2003,Kourakis2005,Chowdhury2018,Fedele2002,Moslem2011}.
It describes not only the stability of DAWs but also the nature of the energy localization
and re-distribution in various plasma medium. Envelope solitonic solutions, which precisely
describe the energy localization and re-distribution, is one of the mysterious solution of the NLSE.
On the other hand, the ambiguous appearance of the rogue waves (RWs) is also an
interesting nonlinear phenomena in inter-disciplinary science (viz., nonlinear fiber optics,
parametrically driven capillary waves, super-fluids, optical cavities, plasmonics \cite{Tolba2015},
hydrodynamics \cite{Benjamin1967}, biology \cite{Turing1952}, Bose-Einstein condensates, ocean waves,
and stock market \cite{Shalini2015}, etc.). The RWs are investigated in a MCPM and have been
experimentally observed and also modeled by using the NLSE \cite{Chabchoub2011,Chabchoub2012,Bailung2011,Kibler2010,Shalini2015}.
Zaghbeer \textit{et al.} \cite{Zaghbeer2014} investigated the MI of the DAWS in
a four components plasma medium with non-extensive electrons and ions, and observed that the variation of the non-extensivity
of electron and ion would lead to increase the amplitude of DARWs. Eghbali \textit{et al.} \cite{Eghbali2017}
numerically analyzed the criterion of the MI of DIAWs in presence of non-extensive electrons in non-planar geometry,
and found that the value of maximum growth rate decreases with increasing $q$. Moslem \textit{et al.} \cite{Moslem2011}
demonstrated three components plasma medium having non-extensive electrons and ions as well as massive negatively charged dust grains
to study the stability of the DAWs, and reported that for negative $q$, the $k_c$ increases with $q$.
Jahan \textit{et al.} \cite{Jahan2019} considered an OPDGs to investigate the MI of the DAWs, and found that
the critical wave number ($k_c$) decreases with increasing the negative dust charge state when other parameters remain constant.
Chowdhury \textit{et al.} \cite{Chowdhury2018} studied the amplitude modulation of the nucleus-acoustic waves, and
observed that the variation of the plasma parameters causes to change the thickness but not the height of the the envelope pulses.

Recently, Hossen \textit{et al.} \cite{Hossen2017} investigated the DASWs in a three components EDDP
having inertial an OPDGs and inertialess $q$-distributed ions. Ferdousi \textit{et al.} \cite{Ferdousi2015} studied DASHWs in a two
components EDDPs in presence of non-extensive ions, and observed that both polarities of DASHWs can
exist according to the value of $q$. This is all very nice and certainly deserves further examination,
the purpose of this article is to extend Ferdousi \textit{et al.} \cite{Ferdousi2015} work by deriving
a NLSE and also investigate the MI of DAWs and associated DA electrostatic envelope solitons as well as first-order and second-order
DARWs in a three components EDDP having inertial warm positively and negatively charged massive
dust grains as well as inertialess non-extensive
$q$-distributed ions.

The outline of the paper is as follows: The governing equations describing our plasma
model are presented in Section \ref{1sec:Governing Equations}. The NLSE is derived in
Section \ref{1sec:Derivation of the NLSE}. Modulational instability is given in Section
\ref{1sec:Modulational instability}. The formation of envelope solitons and rogue waves are, respectively, presented
in Sections \ref{1sec:Envelope solitons} and \ref{1sec:Rogue waves}.
A brief conclusion is finally provided in Section \ref{1sec:Conclusion}.
\section{Governing Equations}
\label{1sec:Governing Equations}
We consider an unmagnetized three components EDDP medium
consisting of inertial warm positively and negatively charged massive
dust grains, and inertialess non-extensive $q$-distributed positive ions. At equilibrium,
the quasi-neutrality condition for our plasma model can be written as $ Z_i n_{i0}+Z_+ n_{+0}\approx Z_- n_{-0}$;
where $n_{i0}$, $n_{+0}$, and $n_{-0}$ are, respectively, the equilibrium number densities of
positive ions, warm positive and negative dust grains, and also  $Z_i$, $Z_+$, and $Z_-$ are, respectively, the charge state of
the positive ions, warm positive and negative dust grains.
Now, the normalized governing equations for studying DAWs can be written as
\begin{eqnarray}
&&\hspace*{-1.3cm}\frac{\partial {n_+}}{\partial t}+\frac{\partial}{\partial {x}}(n_+u_+)=0,
\label{1eq:1}\\
&&\hspace*{-1.3cm}\frac{\partial {u_+}}{\partial t}+u_+\frac{\partial {u_+}}{\partial {x}}+3 e_1n_+\frac{\partial {n_+}}{\partial x}=-\frac{\partial {\phi}}{\partial x},
\label{1eq:2}\\
&&\hspace*{-1.3cm}\frac{\partial {n_-}}{\partial t}+\frac{\partial}{\partial {x}}(n_-u_-)=0,
\label{1eq:3}\\
&&\hspace*{-1.3cm}\frac{\partial {u_-}}{\partial t}+u_-\frac{\partial {u_-}}{\partial {x}}+3 e_2n_-\frac{\partial {n_-}}{\partial x}=e_3\frac{\partial {\phi}}{\partial x},
\label{1eq:4}\\
&&\hspace*{-1.3cm}\frac{\partial^2\phi}{\partial x^2}=e_4n_--(e_4-1)n_i-n_+,
\label{1eq:5}\
\end{eqnarray}
where $n_+$, $n_-$, and $n_i$ are the number densities of positively charged dust grains,
negatively charged dust grains, and positive ions normalized by their equilibrium values
$n_{+0}$, $n_{-0}$, and $n_{i0}$, respectively; $u_+$ and $u_-$ are the dust fluid speeds
normalized by the DA wave speed $C_+=(Z_+k_BT_i/m_+)^{1/2}$ (with $T_i$ being the ion
temperature, $m_+$ being the positive dust mass, and $k_B$ being the Boltzmann constant);
$\phi$ is the electrostatic wave potential normalized by $k_BT_i/e$ (with $e$ being the
magnitude of single electron charge); the time and space variables are normalized by
$\omega_{P+}^{-1}=({m_+}/4\pi e^2Z_+^2n_{+0})^{1/2}$ and $\lambda_{D} = (k_BT_i/4\pi e^2Z_+n_{+0})^{1/2}$,
respectively; $P_+=P_{+0}(N_+/n_{+0})^\gamma$ [with $P_{+0}$ being the equilibrium adiabatic
pressure of the warm positive dust grains and $\gamma=(N+2)/N$, where $N$ is the degree of freedom,
for one dimensional case, $N$=1 so that $\gamma$=3]; $P_{+0}=n_{+0}k_BT_+$ (with $T_+$ being the temperature
of the warm positive dust grains and $k_B$ being the Boltzmann constant);
$P_-=P_{-0}(N_-/n_{-0})^\gamma$ [with $P_{-0}$ being the equilibrium adiabatic
pressure of the warm negative dust grains]; $P_{-0}=n_{-0}k_BT_-$ (with $T_-$ being the temperature
of the warm negative dust grains); and $e_1=T_+/Z_+T_i$, $e_2=T_-m_+/Z_+T_im_-$,
$e_3=Z_-m_+/Z_+m_-$, and $e_4=Z_-n_{-0}/Z_+n_{+0}$. We have considered for our numerical analysis
$T_+>T_-$, $T_i\gg (T_+,~T_-)$, $m_->m_+$, $Z_->Z_+$, and $n_{-0}>n_{+0}$.
Now, the $q$ distributed positive ion number density can be expressed in the following form \cite{Tsallis1988,Zaghbeer2014}
\begin{eqnarray}
&&\hspace*{-1.3cm}n_i=[1-(q-1)\phi]^{\frac{q+1}{2(q-1)}},
\label{1eq:6}\
\end{eqnarray}
where the parameter $q$ is known as entropic index, and is used to explain the effects of non-extensivity
on nonlinear structure of DAWs. It is well known that $q<1$ refers to the super-extensivity while
$q>1$ refers to the sub-extensivity, and $q=1$ indicates the Maxwellian distribution.
By substituting Eq. \eqref{1eq:6} into Eq. \eqref{1eq:5} and expanding up to third order in $\phi$, we get
\begin{eqnarray}
&&\hspace*{-1.3cm}\frac{\partial^2\phi}{\partial x^2}+n_+-e_4n_-=1-e_4+S_1\phi
\nonumber\\
&&\hspace*{1.4cm}+S_2\phi^2+S_3\phi^3+\cdots,
\label{1eq:7}\
\end{eqnarray}
where
\begin{eqnarray}
&&\hspace*{-1.3cm}S_1=[(e_4-1)(q+1)]/2,
\nonumber\\
&&\hspace*{-1.3cm}S_2=[(1-e_4)(q+1)(3-q)]/8,
\nonumber\\
&&\hspace*{-1.3cm}S_3=[(1-e_4)(q+1)(3-q )(3q-5)]/48.
\nonumber\
\end{eqnarray}
We note that the right hand side of the Eq. \eqref{1eq:7} is the contribution of positive ions.
\section{Derivation of the NLSE}
\label{1sec:Derivation of the NLSE}
To study the MI of the DAWs, we want to derive the NLSE by employing the reductive perturbation method (RPM)
and for that purpose, we can write the stretched coordinates in the form \cite{Kourakis2003,Kourakis2005,Chowdhury2018}
\begin{eqnarray}
&&\hspace*{-1.3cm}\xi={\epsilon}(x-v_g t),
\label{1eq:8}\\
&&\hspace*{-1.3cm}\tau={\epsilon}^2 t,
\label{1eq:9}\
\end{eqnarray}
where $v_g$ is the group velocity and $\epsilon$ ($\epsilon \ll 1$) is a small parameter.
Then, we can write the dependent variables  as \cite{Kourakis2003,Kourakis2005,Chowdhury2018}
\begin{eqnarray}
&&\hspace*{-1.3cm}n_+=1+\sum_{m=1}^{\infty}\epsilon^{m}\sum_{l=-\infty}^{\infty}n_{+l}^{(m)}(\xi,\tau)~\mbox{exp}[i l(kx-\omega t)],
\label{1eq:10}\\
&&\hspace*{-1.3cm}u_+=\sum_{m=1}^{\infty}\epsilon^{m}\sum_{l=-\infty}^{\infty}u_{+l}^{(m)}(\xi,\tau)~\mbox{exp}[i l(kx-\omega t)],
\label{1eq:11}\\
&&\hspace*{-1.3cm}n_-=1+\sum_{m=1}^{\infty}\epsilon^{m}\sum_{l=-\infty}^{\infty}n_{-l}^{(m)}(\xi,\tau)~\mbox{exp}[i l(kx-\omega t)],
\label{1eq:12}\\
&&\hspace*{-1.3cm}u_-=\sum_{m=1}^{\infty}\epsilon^{m}\sum_{l=-\infty}^{\infty}u_{-l}^{(m)}(\xi,\tau)~\mbox{exp}[i l(kx-\omega t)],
\label{1eq:13}\\
&&\hspace*{-1.3cm}\phi=\sum_{m=1}^{\infty}\epsilon^{m}\sum_{l=-\infty}^{\infty}\phi^{(m)}(\xi,\tau)~\mbox{exp}[i l(kx-\omega t)],
\label{1eq:14}\
\end{eqnarray}
where $k$ and $\omega$ are real variables representing the carrier wave number and frequency, respectively.
The derivative operators in the above equations are treated as follows:
\begin{eqnarray}
&&\hspace*{-1.3cm}\frac{\partial}{\partial t}\rightarrow\frac{\partial}{\partial t}-\epsilon v_g \frac{\partial}{\partial\xi}+\epsilon^2\frac{\partial}{\partial\tau},
\label{1eq:15}\\
&&\hspace*{-1.3cm}\frac{\partial}{\partial x}\rightarrow\frac{\partial}{\partial x}+\epsilon\frac{\partial}{\partial\xi}.
\label{1eq:16}
\end{eqnarray}
Now, by substituting \eqref{1eq:8}-\eqref{1eq:16}  into  \eqref{1eq:1}-\eqref{1eq:4}, and \eqref{1eq:7}, and
equating the coefficients of $\epsilon$ for $m=l=1$, then we get the following equations
\begin{eqnarray}
&&\hspace*{-1.3cm}k u_{+1}^{(1)}=\omega n_{+1}^{(1)},
\label{1eq:17}\\
&&\hspace*{-1.3cm}k \phi_{1}^{(1)}+k\alpha n_{+1}^{(1)}=\omega n_{+1}^{(1)},
\label{1eq:18}\\
&&\hspace*{-1.3cm}k u_{-1}^{(1)}=\omega n_{-1}^{(1)},
\label{1eq:19}\\
&&\hspace*{-1.3cm}k\beta n_{-1}^{(1)}=\omega n_{-1}^{(1)}+ke_3 \phi_{1}^{(1)},
\label{1eq:20}\\
&&\hspace*{-1.3cm}n_{+1}^{(1)}=k^2\phi_{1}^{(1)}+S_1\phi_{1}^{(1)}+e_4n_{-1}^{(1)},
\label{1eq:21}
\end{eqnarray}
where $\alpha=3e_1$ and $\beta=3e_2$. These equations reduce to
\begin{eqnarray}
&&\hspace*{-1.3cm}n_{+1}^{(1)}=\frac{k^2}{A}\phi^{(1)}_1,
\label{1eq:22}\\
&&\hspace*{-1.3cm}u_{+1}^{(1)}=\frac{\omega k}{A}\phi^{(1)}_1,
\label{1eq:23}\\
&&\hspace*{-1.3cm}n_{-1}^{(1)}=\frac{e_3k^2}{C}\phi^{(1)}_1,
\label{1eq:24}\\
&&\hspace*{-1.3cm}u_{-1}^{(1)}=\frac{\omega k e_3}{C}\phi^{(1)}_1,
\label{1eq:25}\
\end{eqnarray}
where $A=\omega^2-\alpha k^2$ and $C=\beta k^2-\omega^2$. Therefore, the dispersion relation can be written as
\begin{eqnarray}
&&\hspace*{-1.3cm}\omega^2=\frac{k^2M\pm k^2\sqrt{M^2-4GH}}{2G},
\label{1eq:26}\
\end{eqnarray}
\begin{figure}
\centering
\includegraphics[width=80mm]{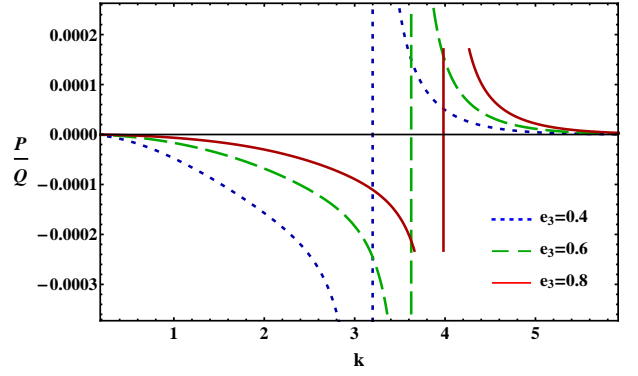}
\caption{Plot of $P/Q$ vs $k$ for different values of $e_3$ when $e_1=0.07$, $e_2=0.007$, $e_4=2.0$, $q=2$, and $\omega_s$.}
\label{1Fig:F1}
\end{figure}
where $M=1+e_3e_4+\alpha k^2+\alpha S_1+\beta k^2+\beta S_1$, $G=k^2+S_1$, and $H=\beta+\alpha\beta k^2+\alpha\beta S_1+\alpha e_3e_4$.
In order to obtain the real positive values of $\omega$, the condition $M^2>4GH$ must be satisfied.
The positive and negative signs of Eq. \eqref{1eq:26} determine two types of DA modes, i.e., the positive
sign refers to the fast ($\omega_f$) DA mode whereas the negative sign refers to the slow
($\omega_s$) DA mode. The fast DA mode corresponds to the case in which both
dust species oscillate in phase with ions. On
the other hand, the slow DA mode corresponds to the case
in which only one of the inertial massive dust components
oscillates in phase with ions but the other species
are in anti-phase with them \cite{Jahan2019}.
The second-order ($m=2$ with $l=1$) equations are given by
\begin{eqnarray}
&&\hspace*{-1.3cm}n_{+1}^{(2)}=\frac{k^2}{A}\phi_1^{(2)}+\frac{ikB_1}{A^2} \frac{\partial \phi_1^{(1)}}{\partial\xi},
\label{1eq:27}\\
&&\hspace*{-1.3cm}u_{+1}^{(2)}=\frac{\omega k}{A}\phi_1^{(2)}+\frac{iB_2}{A^2} \frac{\partial \phi_1^{(1)}}{\partial\xi},
\label{1eq:28}\\
&&\hspace*{-1.3cm}n_{-1}^{(2)}=\frac{e_3k^2}{C}\phi_1^{(2)}-\frac{ike_3B_3}{C^2} \frac{\partial \phi_1^{(1)}}{\partial\xi},
\label{1eq:29}\\
&&\hspace*{-1.3cm}u_{-1}^{(2)}=\frac{\omega k e_3}{C}\phi_1^{(2)}-\frac{ie_3B_4}{C^2} \frac{\partial \phi_1^{(1)}}{\partial\xi},
\label{1eq:30}\
\end{eqnarray}
where
\begin{eqnarray}
&&\hspace*{-1.3cm}B_1=2\omega k v_g-\omega^2-A-\alpha k^2,
\nonumber\\
&&\hspace*{-1.3cm}B_2=2kv_g\omega^2-\omega^3-\alpha\omega k^2-kv_gA,
\nonumber\\
&&\hspace*{-1.3cm}B_3=2\omega k v_g-\omega^2+C-\beta k^2,
\nonumber\\
&&\hspace*{-1.3cm}B_4=2kv_g\omega^2 -\omega^3-\beta\omega k^2+kv_gC,
\nonumber\
\end{eqnarray}
and the group velocity of DAWs, with the compatibility condition, can be written as
\begin{eqnarray}
&&\hspace*{-1.3cm}v_g=\frac{\partial \omega}{\partial k}=\frac{B_5-2A^2C^2}{2\omega k(C^2+e_3e_4A^2)},
\label{1eq:31}\
\end{eqnarray}
where $B_5=e_3e_4A^2\omega^2-e_3e_4CA^2+e_3e_4\beta A^2k^2+\alpha k^2C^2+AC^2+\omega^2C^2$.
The coefficients of $\epsilon$ for $m=2$ and $l=2$ provide the second-order harmonic amplitudes
which are found to be proportional to $|\phi^{(1)}_1|^2$
\begin{eqnarray}
&&\hspace*{-1.3cm}n_{+2}^{(2)}=S_4|\phi_1^{(1)}|^2,
\label{1eq:32}\\
&&\hspace*{-1.3cm}u_{+2}^{(2)}=S_5 |\phi_1^{(1)}|^2,
\label{1eq:33}\\
&&\hspace*{-1.3cm}n_{-2}^{(2)}=S_6|\phi_1^{(1)}|^2,
\label{1eq:34}\\
&&\hspace*{-1.3cm}u_{-2}^{(2)}=S_7 |\phi_1^{(1)}|^2,
\label{1eq:35}\\
&&\hspace*{-1.3cm}\phi_{2}^{(2)}=S_8 |\phi_1^{(1)}|^2,
\label{1eq:36}\
\end{eqnarray}
where
\begin{eqnarray}
&&\hspace*{-1.3cm}S_4=\frac{2k^2A^2S_8+\alpha k^6+3\omega^2k^4}{2A^3},
\nonumber\\
&&\hspace*{-1.3cm}S_5=\frac{\omega S_4A^2-\omega k^4}{kA^2},
\nonumber\\
&&\hspace*{-1.3cm}S_6=\frac{2e_3k^2C^2S_8-\beta e_3^2k^6-3\omega^2 k^4}{2C^3},
\nonumber\\
&&\hspace*{-1.3cm}S_7=\frac{\omega S_6C^2-\omega e_3^2k^4}{kC^2},
\nonumber\\
&&\hspace*{-1.3cm}S_8=\frac{B_6}{B_7},
\nonumber\\
&&\hspace*{-1.3cm}B_6=2S_2A^3C^3-\beta e_4e_3^2A^3k^6-3e_4e_3^2\omega^2k^4A^3
\nonumber\\
&&\hspace*{-0.5cm}-\alpha C^3 k^6-3\omega^2k^4C^3,
\nonumber\\
&&\hspace*{-1.3cm}B_7=2A^2k^2C^3-2e_3e_4A^3C^2k^2-8k^2A^3C^3
\nonumber\\
&&\hspace*{-0.5cm}-2S_1A^3C^3.
\nonumber\
\end{eqnarray}
Now, we consider the expression for ($m=3$ with $l=0$) and ($m=2$ with $l=0$),
which leads to the zeroth harmonic modes. Thus, we obtain
\begin{eqnarray}
&&\hspace*{-1.3cm}n_{+0}^{(2)}=S_9|\phi_1^{(1)}|^2,
\label{1eq:37}\\
&&\hspace*{-1.3cm}u_{+0}^{(2)}=S_{10}|\phi_1^{(1)}|^2,
\label{1eq:38}\\
&&\hspace*{-1.3cm}n_{-0}^{(2)}=S_{11}|\phi_1^{(1)}|^2,
\label{1eq:39}\\
&&\hspace*{-1.3cm}u_{-0}^{(2)}=S_{12}|\phi_1^{(1)}|^2,
\label{1eq:40}\\
&&\hspace*{-1.3cm}\phi_0^{(2)}=S_{13}|\phi_1^{(1)}|^2,
\label{1eq:41}\
\end{eqnarray}
where
\begin{eqnarray}
&&\hspace*{-1.3cm}S_9=\frac{2\omega v_gk^3+\alpha k^4+k^2\omega^2+S_{13}A^2}{A^2(v_g^2-\alpha)},
\nonumber\\
&&\hspace*{-1.3cm}S_{10}=\frac{v_gS_9A^2-2\omega k^3}{A^2},
\nonumber\\
&&\hspace*{-1.3cm}S_{11}=\frac{\beta e_3^2k^4+2\omega v_ge_3^2k^3+e_3^2k^2\omega^2-e_3S_{13}C^2}{C^2(v_g^2-\beta)},
\nonumber\\
&&\hspace*{-1.3cm}S_{12}=\frac{v_gS_{11}C^2-2\omega e_3^2k^3}{C^2},
\nonumber\\
&&\hspace*{-1.3cm}S_{13}=\frac{2S_2A^2C^2(v_g^2-\alpha)(v_g^2-\beta)+B_8}{B_9A^2C^2+A^2C^2(v_g^2-\beta)},
\nonumber\\
&&\hspace*{-1.3cm}B_8=e_4e_3^2A^2(v_g^2-\alpha)(\beta k^4+2\omega v_gk^3+k^2\omega^2)
\nonumber\\
&&\hspace*{-0.5cm}-C^2(v_g^2-\beta)(\alpha k^4+2\omega v_gk^3+k^2\omega^2),
\nonumber\\
&&\hspace*{-1.3cm}B_9=e_4e_3(v_g^2-\alpha)-S_1(v_g^2-\alpha)(v_g^2-\beta).
\nonumber\
\end{eqnarray}
Finally, the third harmonic modes ($m=3$) and ($l=1$) gives a set of equations, which can be reduced
to the following NLSE:
\begin{eqnarray}
&&\hspace*{-1.3cm}i\frac{\partial\Phi}{\partial\tau}+P\frac{\partial^2\Phi}{\partial\xi^2}+Q|\Phi|^2\Phi=0,
\label{1eq:42}
\end{eqnarray}
where $\Phi=\phi_1^{(1)}$ for simplicity. In equation \eqref{1eq:42}, $P$ is the dispersion
coefficient which can be written as
\begin{eqnarray}
&&\hspace*{-1.3cm}P=\frac{B_{10}-A^3C^3}{2\omega k^2AC(C^2+e_3e_4A^2)},
\nonumber\
\end{eqnarray}
where
\begin{eqnarray}
&&\hspace*{-1.3cm}B_{10}=e_3e_4A^3(4\beta\omega v_gk^3+4v_gk\omega^3+\beta k^2C+k^2v_g^2C
\nonumber\\
&&\hspace*{-0.5cm}-4k^2\omega^2v_g^2-2\beta k^2\omega^2-\beta^2k^4-\omega^4)
\nonumber\\
&&\hspace*{-0.5cm}-C^3(4\alpha\omega v_gk^3+4v_gk\omega^3+\alpha k^2A+k^2v_g^2A
\nonumber\\
&&\hspace*{-0.5cm}-4k^2\omega^2v_g^2-2\alpha k^2\omega^2-\alpha^2k^4-\omega^4),
\nonumber\
\end{eqnarray}
and $Q$ is the nonlinear coefficient which can be written as
\begin{eqnarray}
&&\hspace*{-1.3cm}Q=\frac{2S_2A^2C^2(S_8+S_{13})+3S_3A^2C^2-B_{11}}{2\omega k^2(C^2+e_3e_4A^2)},
\nonumber\
\end{eqnarray}
where
\begin{eqnarray}
&&\hspace*{-1.3cm}B_{11}=\alpha k^4S_4C^2+\alpha k^4S_9C^2+\beta e_3e_4k^4S_6A^2
\nonumber\\
&&\hspace*{-0.5cm}+2e_3e_4\omega k^3S_{12}A^2+k^2\omega^2S_4C^2+k^2\omega^2S_9C^2
\nonumber\\
&&\hspace*{-0.5cm}+2\omega S_5k^3C^2+2\omega k^3S_{10}C^2+2e_3e_4\omega k^3S_7A^2
\nonumber\\
&&\hspace*{-0.5cm}+\beta e_3e_4k^4S_{11}A^2+e_3e_4S_6k^2\omega^2+e_3e_4S_{11}k^2\omega^2.
\nonumber\
\end{eqnarray}
It may be noted here that both $P$ and $Q$ are function of various
plasma parameters such as  $q$, $e_1$, $e_2$, $e_3$, $e_4$ and $k$.
So, all the plasma parameters are used to maintain
the nonlinearity and the dispersion properties of the EDDP medium.
\begin{figure}
\centering
\includegraphics[width=80mm]{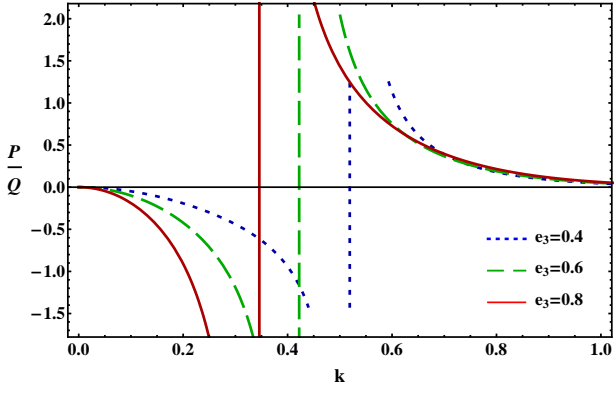}
\caption{Plot of $P/Q$ vs $k$ for  different values of $e_3$ when $e_1=0.07$, $e_2=0.007$, $e_4=2.0$, $q=1$, and $\omega_f$.}
 \label{1Fig:F2}
\end{figure}
\begin{figure}
\centering
\includegraphics[width=80mm]{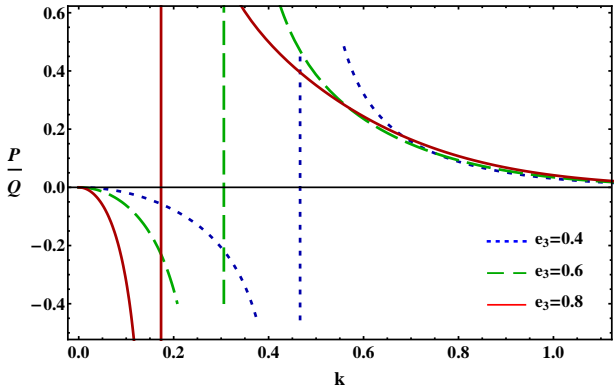}
\caption{Plot of $P/Q$ vs $k$ for  different values of $e_3$ when $e_1=0.07$, $e_2=0.007$, $e_4=2.0$, $q=2$, and $\omega_f$.}
 \label{1Fig:F3}
\end{figure}
\section{Modulational instability}
\label{1sec:Modulational instability}
The stable and unstable parametric regimes of the DAWs are organized by the sign of the dispersion ($P$)
and nonlinear ($Q$) coefficients of the standard NLSE \eqref{1eq:42} \cite{Kourakis2003,Kourakis2005,Chowdhury2018,Fedele2002,Moslem2011,C4,C5,C6}.
When $P$ and $Q$ have same sign (i.e., $P/Q>0$), the evolution of the DAWs amplitude is
modulationally unstable in presence of the external perturbations. On the other hand,
when $P$ and $Q$ have opposite sign (i.e., $P/Q<0$), the DAWs are modulationally stable.
The plot of $P/Q$ against $k$ yields stable and unstable parametric regimes for the DAWs.
The point, at which transition of $P/Q$ curve intersects with $k$-axis, is known as threshold
or critical wave number $k$ ($=k_c$).

Figure \ref{1Fig:F1} shows the variation of $P/Q$ with $k$ for
different values of $e_3$ and also for DA slow mode ($\omega_s$). It is clear from this figure
that (a) the stable parametric regime of DAWs increases with increasing (decreasing) the mass of the
positive (negative) dust grains while $Z_+$ and $Z_-$ remain constant; (b) an increase in the
value of negative dust charge state causes to increase the stable parametric regime of DAWs
whereas an increase in the value of positive dust charge state  causes to increase the unstable
parametric regime for a fixed value of $m_+$ and $m_-$ (via  $e_3$).

\begin{figure}
\centering
\includegraphics[width=80mm]{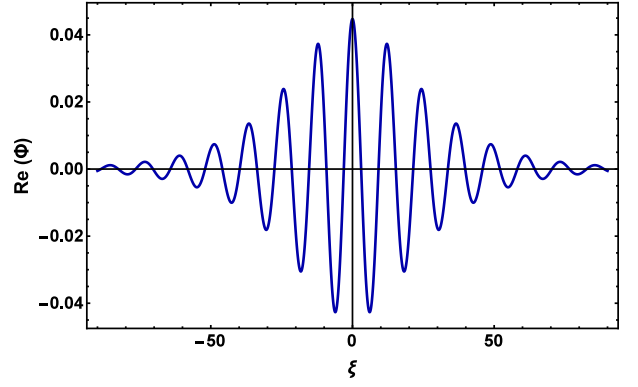}
\caption{Plot of $\mbox{Re}(\Phi)$ vs $\xi$ for bright envelope soliton when $e_1=0.07$, $e_2=0.007$, $e_3=0.6$, $e_4=2.0$, $q=2$, $\tau=0$, $\psi_0=0.002$, $U=0.4$, $\Omega_0=0.4$, $k=0.5$, and $\omega_f$.}
 \label{1Fig:F4}
\end{figure}
\begin{figure}
\centering
\includegraphics[width=80mm]{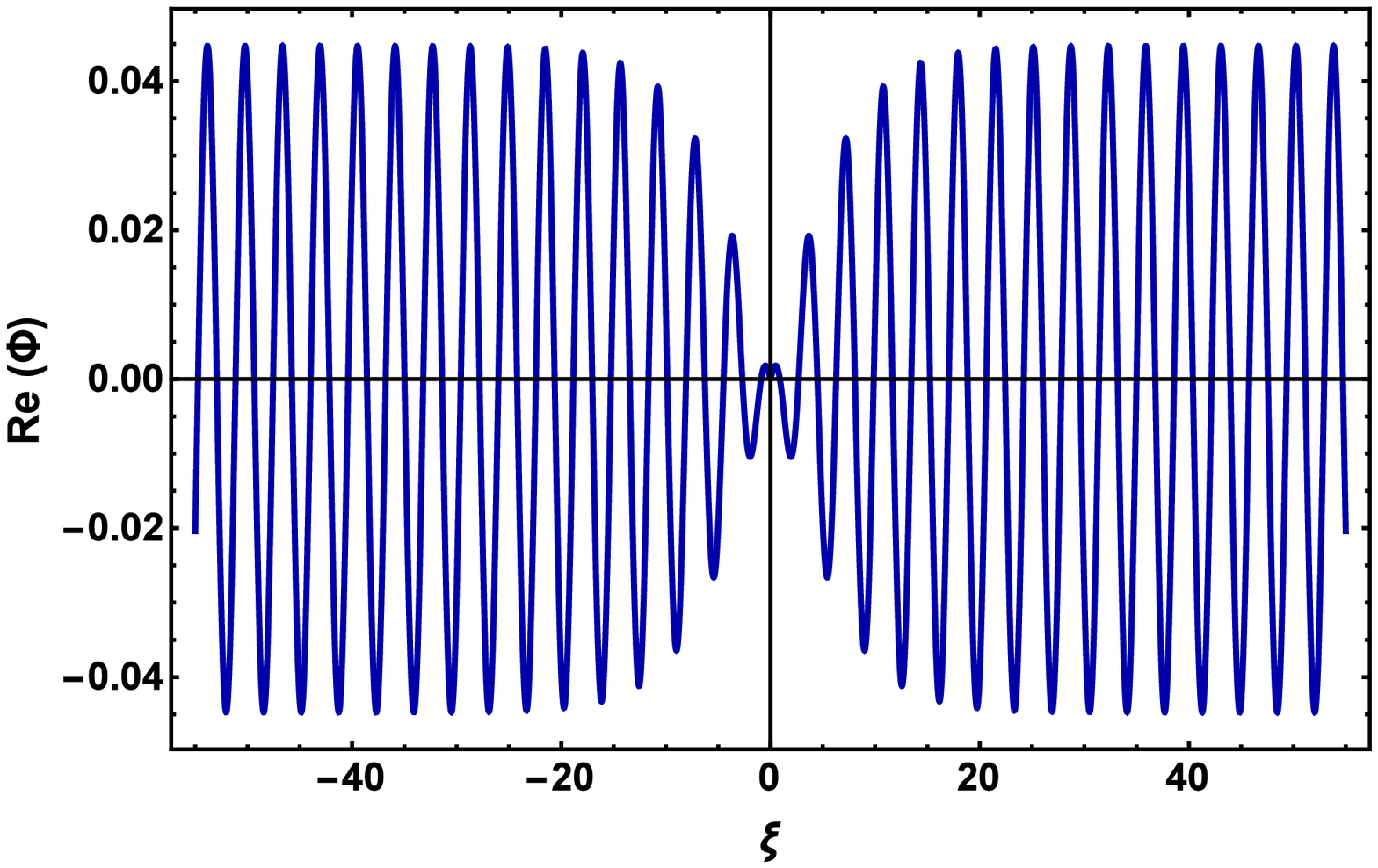}
\caption{Plot of $\mbox{Re}(\Phi)$ vs $\xi$ for dark envelope soliton when $e_1=0.07$, $e_2=0.007$, $e_3=0.6$, $e_4=2.0$, $q=2$, $\tau=0$, $\psi_0=0.002$, $U=0.4$, $\Omega_0=0.4$, $k=0.1$, and $\omega_f$.}
\label{1Fig:F5}
\end{figure}
\begin{figure}
\centering
\includegraphics[width=80mm]{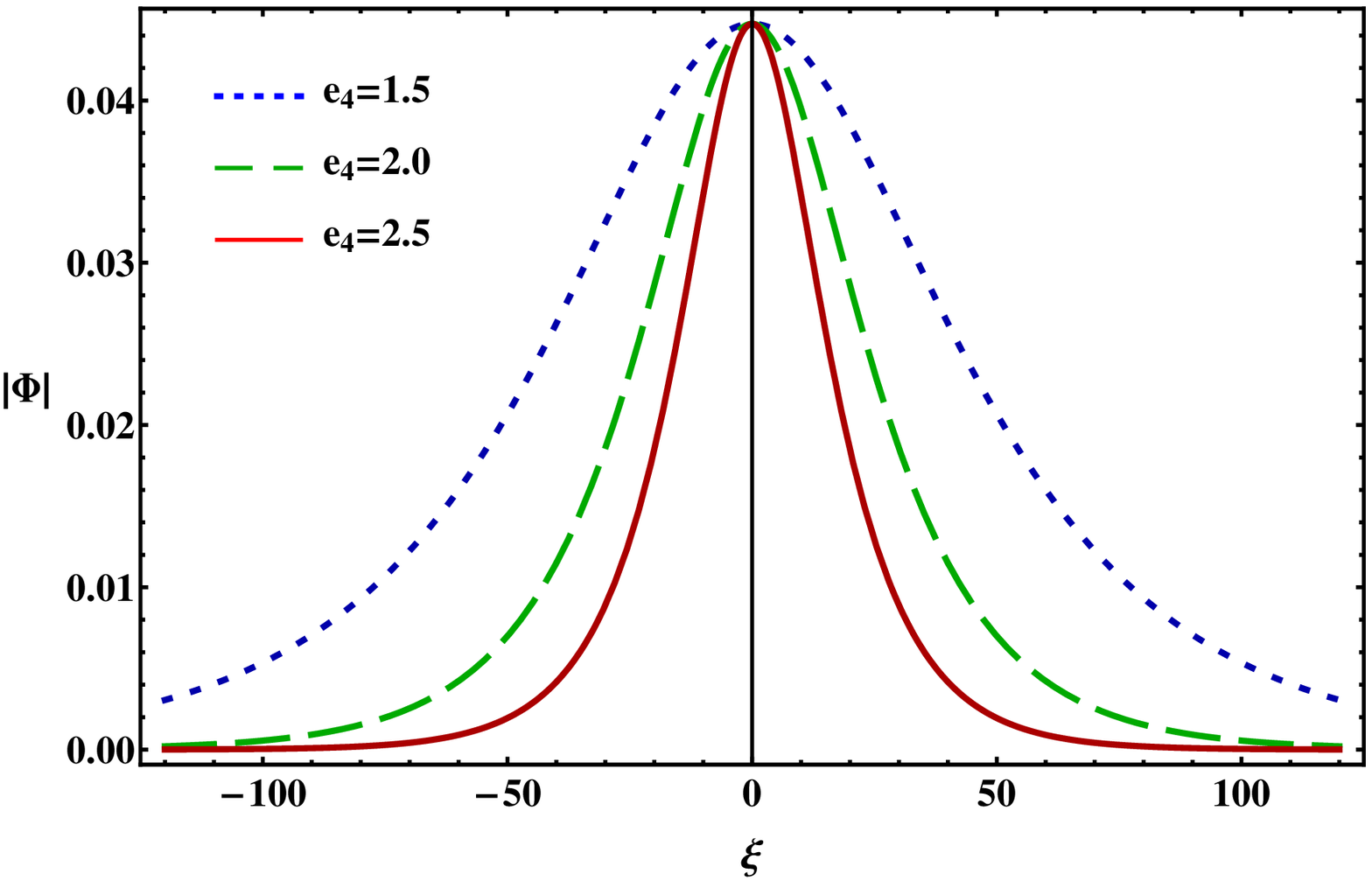}
\caption{Plot of $|\Phi|$ vs $\xi$ for  different values of $e_4$ when $e_1=0.07$, $e_2=0.007$, $e_3=0.6$, $q=2$, $\tau=0$, $\psi_0=0.002$, $U=0.4$, $\Omega_0=0.4$, $k=0.5$, and $\omega_f$.}
 \label{1Fig:F6}
\end{figure}
\begin{figure}[t!]
\centering
\includegraphics[width=80mm]{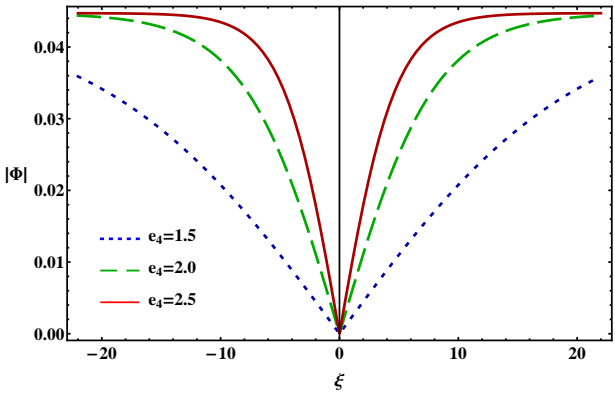}
\caption{Plot of $|\Phi|$ vs $\xi$ for  different values of $e_4$ when $e_1=0.07$, $e_2=0.007$, $e_3=0.6$, $q=2$, $\tau=0$, $\psi_0=0.002$, $U=0.4$, $\Omega_0=0.4$, $k=0.1$, and $\omega_f$.}
\label{1Fig:F7}
\end{figure}
The criteria for the formation of dark envelope solitons associated with stable region (i.e., $P/Q<0$) of DAWs as well as
bright envelope solitons and DARWs associated with unstable region (i.e., $P/Q>0$) of DAWs is shown in Fig. \ref{1Fig:F2} by depicting the variation
of $P/Q$ with $k$ for the values of $e_3$ and for Maxwellian ions (i.e., $q=1$).
It is obvious from this figure that (a) the modulationally stable parametric regime of DAWs reduces (enhances)
as we increase the value of positive (negative) dust mass for a fixed value of $Z_-$ and $Z_+$;
(b) the modulationally unstable parametric regime increases (decreases) with an increase of $Z_-$ ($Z_+$)
for the fixed values of positive and negative dust mass (via $e_3$).

Figure \ref{1Fig:F3} indicates that how the mass and charge state of the positive and negative dust grains of a three components EDDP
can organize the stability criterion for the DAWs. The $k_c$ as well as the stable parametric regime (i.e., $P/Q<0$)
of DAWs decreases with an increase in the value of $e_3$  for a sub-extensive limit of ions (i.e., $q=2$).
It can be deduced from Figs. \ref{1Fig:F2}-\ref{1Fig:F3} that the direction of the
variation of $k_c$ is independent to the two limits of $q$ (i.e., $q>0$ and $q<0$) but
dependent to the variation of the mass and charge state of the positive and negative dust grains.
\section{Envelope solitons}
\label{1sec:Envelope solitons}
The bright and dark envelope solitons can be written as \cite{Kourakis2003,Kourakis2005,Chowdhury2018,Fedele2002}
\begin{eqnarray}
&&\hspace*{-1.3cm}\Phi(\xi,\tau)=\left[\psi_0~\mbox{sech}^2 \left(\frac{\xi-U\tau}{W}\right)\right]^\frac{1}{2}
\nonumber\\
&&\hspace*{0.0cm}\times \exp \left[\frac{i}{2P}\left\{U\xi+\left(\Omega_0-\frac{U^2}{2}\right)\tau \right\}\right],
\label{1eq:43}\\
&&\hspace*{-1.3cm}\Phi(\xi,\tau)=\left[\psi_0~\mbox{tanh}^2 \left(\frac{\xi-U\tau}{W}\right)\right]^\frac{1}{2}
\nonumber\\
&&\hspace*{0.0cm}\times \exp \left[\frac{i}{2P}\left\{U\xi-\left(\frac{U^2}{2}-2 P Q \psi_0\right)\tau \right\}\right],
\label{1eq:44}\
\end{eqnarray}
where $\psi_0$ indicates the envelope amplitude, $U$ is the travelling speed of the localized pulse,
$W$ is the pulse width which can be written as $W=(2P\psi_0/Q)^{1/2}$, and $\Omega_0$ is the
oscillating frequency at $U=0$. We have depicted bright envelope
solitons in Figs. \ref{1Fig:F4} and \ref{1Fig:F6} by using Eq. \eqref{1eq:43}, and also
dark envelope solitons in Figs. \ref{1Fig:F5} and \ref{1Fig:F7} by using Eq. \eqref{1eq:44}.
The characteristics of bright envelope solitons can be observed from Fig. \ref{1Fig:F6}, and
it is clear from this figure that (a) the width of the bright envelope solitions
associated with DAWs enhances with decreasing (increasing) number density of negative (positive) dust when the
charge state of both dust are remain invariant (via $e_4$); (b) the height of the bright
envelope soliton remains constant with the variation of different plasma parameters such as the number density and charge
state of OPDGs of an EDDP, and this result is a good agreement with the work of Chowdhury \textit{et al.} \cite{Chowdhury2018}.

Figure \ref{1Fig:F7} illustrates the effects of the negative dust number density ($n_{-0}$)
and positive dust number density ($n_{+0}$) as well as the number of charges residing on dust grains on the
formation of dark envelope solitons associated with DAWs, and it is obvious from this figure
that (a) the increase in the value of $e_4$ causes to change the width of the dark envelope solitons but does not cause any change
in the magnitude of the amplitude of the dark envelope solitons; (b) the magnitude of the amplitude of
the dark envelope solitons does not depend on any plasma parameters such as number density and charge state
of positive and negative dust grains (via $e_4$), and this result is a good agreement with the work of Chowdhury \textit{et al.} \cite{Chowdhury2018}.
\section{Rogue waves}
\label{1sec:Rogue waves}
The NLSE \eqref{1eq:42} has a variety of solutions,
among them there is a hierarchy of rational solutions that are located
on a nonzero background and localized in both the $\xi$ and $\tau$ variables. Each solution of the
hierarchy represents a unique event in space and time, as it
increases its amplitude quickly along each variable, reaching
its maximum value and finally decays, just as quickly as it
appeared \cite{Ankiewiez2009a}. Thus, these waves were nicknamed ``waves that
appear from nowhere and disappear without a trace \cite{Ankiewiez2009b}''.
The first-order rational solution of NLSE \eqref{1eq:42} is given as \cite{Guo2013a,Ankiewiez2009a,Ankiewiez2009b}
\begin{eqnarray}
&&\hspace*{-1.3cm}\Phi_1 (\xi, \tau)=\sqrt{\frac{2P}{Q}}\Big[\frac{4+16 i\tau P}{1+4 \xi^2 + 16\tau^2 P^2}-1\Big] \mbox{exp} (2i\tau P).
\label{1eq:45}
\end{eqnarray}

\begin{figure}[t!]
\centering
\includegraphics[width=80mm]{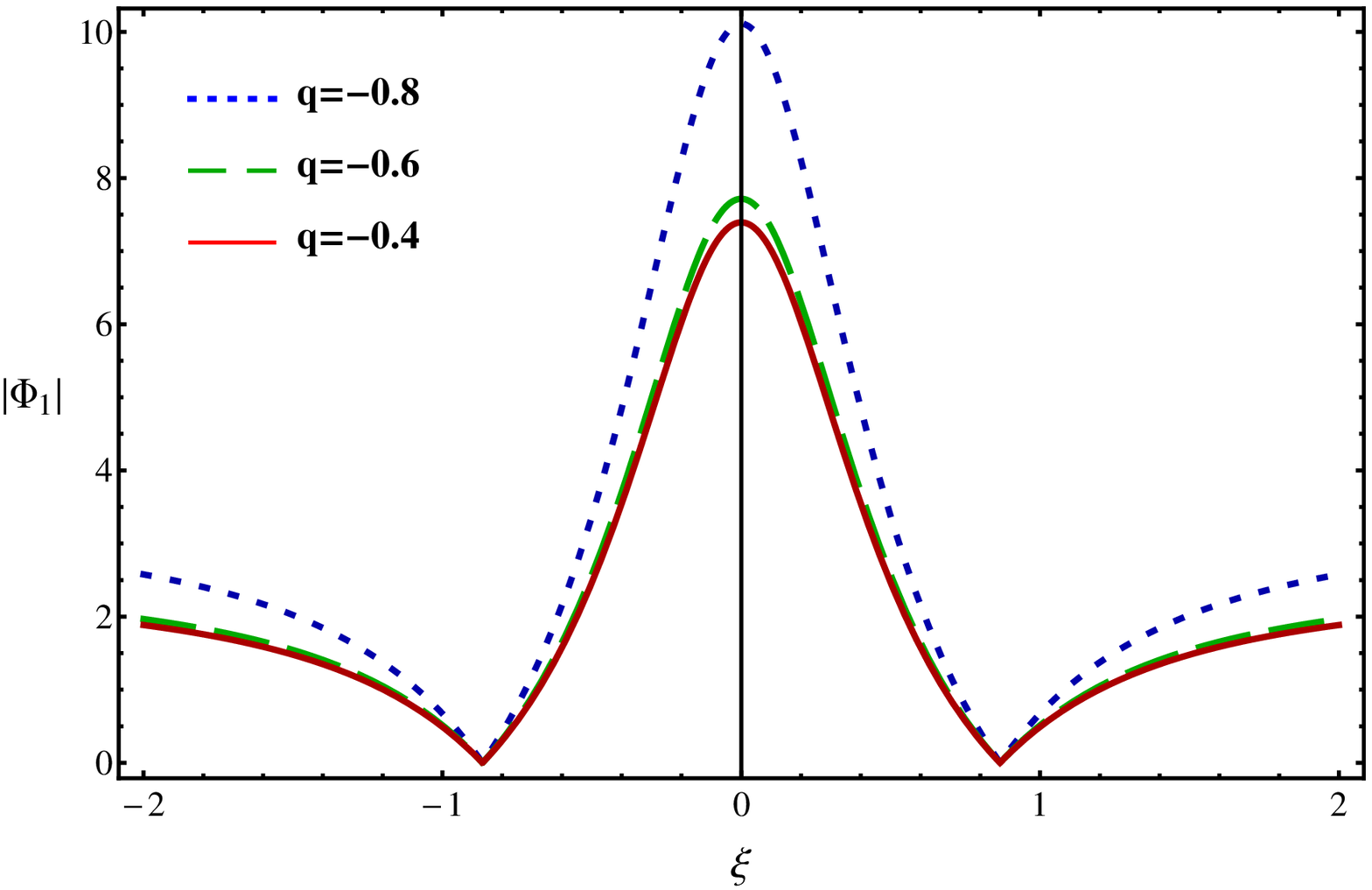}
\caption{Plot of $|\Phi_1|$ vs $\xi$ for different values of $q$ when $e_1=0.07$, $e_2=0.007$, $e_3=0.6$, $e_4=2.0$, $\tau=0$, $k=0.5$, and $\omega_f$.}
\label{1Fig:F8}
\end{figure}

\begin{figure}[t!]
\centering
\includegraphics[width=80mm]{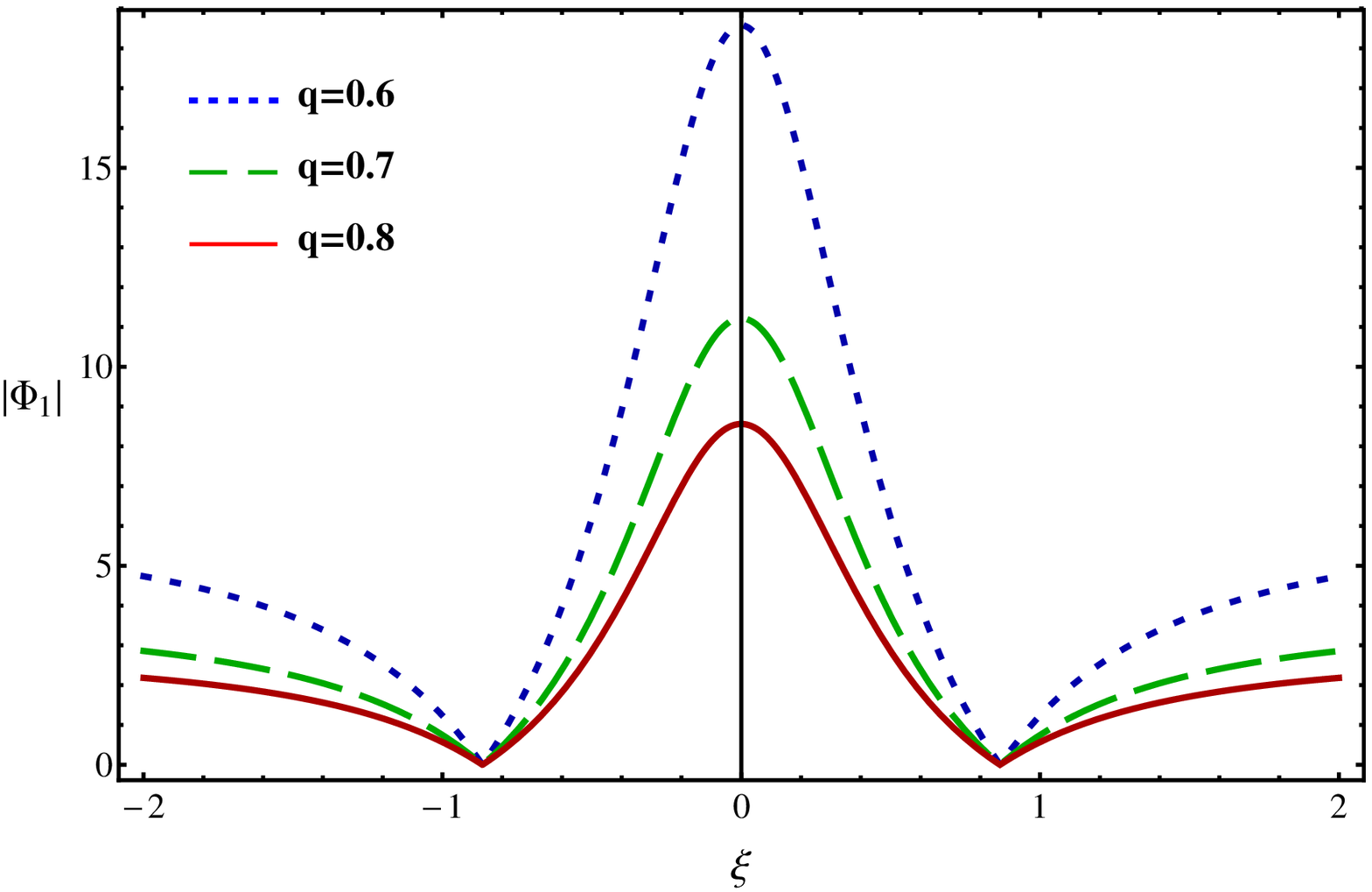}
\caption{Plot of $|\Phi_1|$ vs $\xi$ for different values of $q$ when $e_1=0.07$, $e_2=0.007$, $e_3=0.6$, $e_4=2.0$, $\tau=0$, $k=0.5$, and $\omega_f$.}
\label{1Fig:F9}
\end{figure}

\begin{figure}[t!]
\centering
\includegraphics[width=80mm]{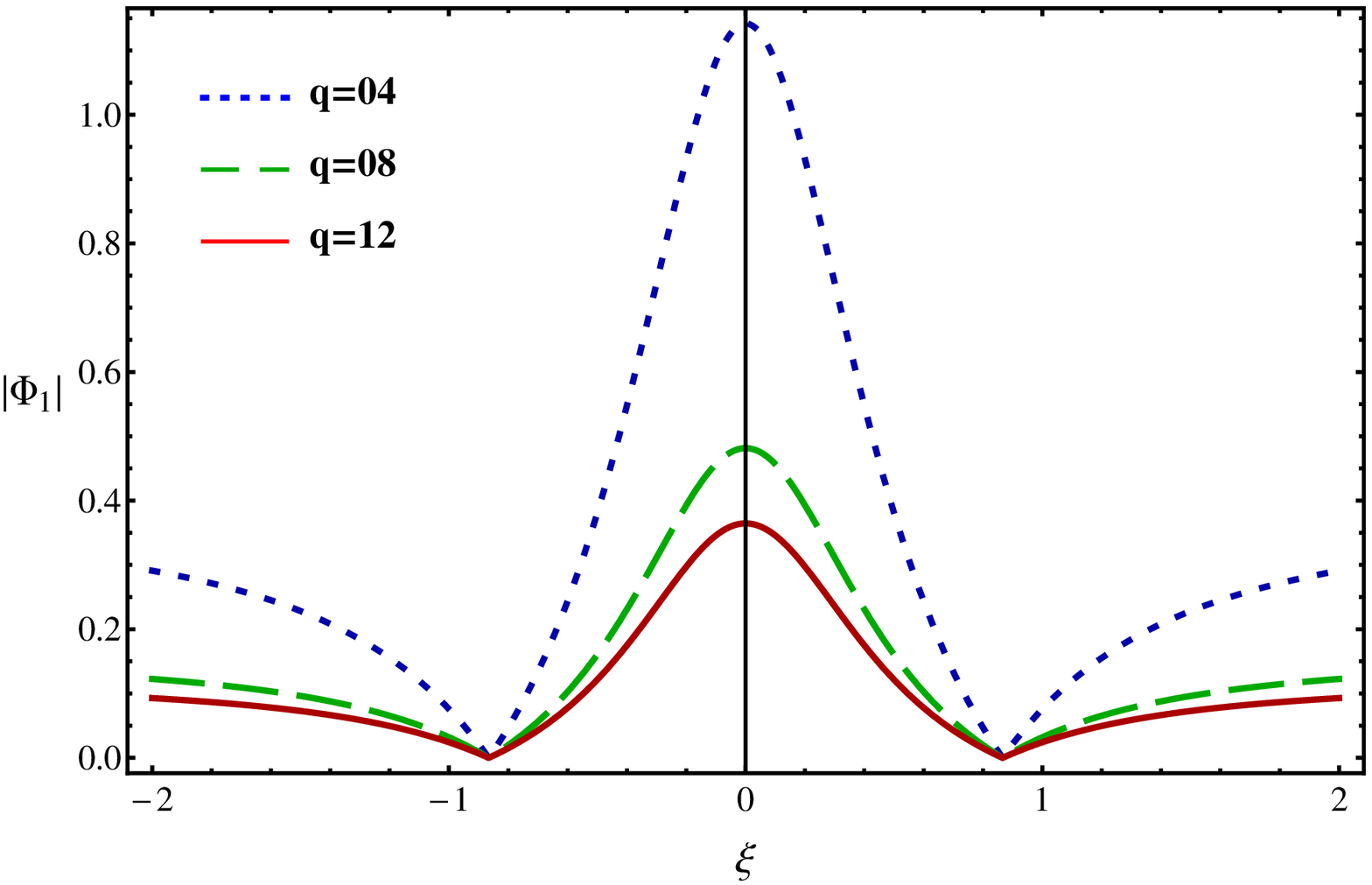}
\caption{Plot of $|\Phi_1|$ vs $\xi$ for different values of $q$ when $e_1=0.07$, $e_2=0.007$, $e_3=0.6$, $e_4=2.0$, $\tau=0$, $k=0.5$, and $\omega_f$.}
\label{1Fig:F10}
\end{figure}

The nonlinear superposition of the two or more first-order RWs gives rise
higher-order RWs and form a more complicated nonlinear structure with high amplitude.
The second-order rational solution is expressed as \cite{Guo2013a,Ankiewiez2009a,Ankiewiez2009b}
\begin{eqnarray}
&&\hspace*{-1.3cm}\Phi_2 (\xi, \tau)=\sqrt{\frac{P}{Q}}\Big[1+\frac{G_2(\xi,\tau)+iM_2(\xi,\tau)}{D_2(\xi,\tau)}\Big] \mbox{exp} (i\tau P).
\label{1eq:46}\
\end{eqnarray}
where
\begin{eqnarray}
&&\hspace*{-1.3cm}G_2(\xi,\tau)=\frac{-\xi^4}{2}-6(P\xi\tau)^2-10(P\tau)^4
\nonumber\\
&&\hspace*{0.2cm}-\frac{3\xi^2}{2}-9(P\tau)^2+\frac{3}{8},
\nonumber\\
&&\hspace*{-1.3cm}M_2(\xi,\tau)=-P\tau\Big[\xi^4+4(P\xi\tau)^2+4(P\tau)^4
\nonumber\\
&&\hspace*{0.2cm}-3\xi^2+2(P\tau)^2-\frac{15}{4}\Big],
\nonumber\\
&&\hspace*{-1.3cm}D_2(\xi,\tau)=\frac{\xi^6}{12}+\frac{\xi^4(P\tau)^2}{2}+\xi^2(P\tau)^4
\nonumber\\
&&\hspace*{0.2cm}+\frac{\xi^4}{8}+\frac{9(P\tau)^4}{2}-\frac{3(P\xi\tau)^2}{2}
\nonumber\\
&&\hspace*{0.2cm}+\frac{9\xi^2}{16}+\frac{33(P\tau)^2}{8}+\frac{3}{32}.
\nonumber\
\end{eqnarray}
The solutions \eqref{1eq:45} and \eqref{1eq:46} represent the profile of the first-order and second-order RWs,
which concentrate a significant amount of energy into a relatively small area, within the modulationally unstable parametric regime.
We have numerically analyzed Eq. \eqref{1eq:45} in Figs. \ref{1Fig:F8}-\ref{1Fig:F10} to understand the effects of
non-extensivity of inertialess ions on the shape of first-order DARWs associated with DAWs
in the modulationally unstable parametric regime, and it is clear from these figures that the nonlinearity as well as the
height and thickness of the first-order DARWs decrease with $q$ for three possible
ranges of $q$ (i.e., $q=$ negative, $q=$ positive but less than 1, and
$q=$ positive but grater than 1). It may be noted here that a number of
authors \cite{Sahu2012a,Ferdousi2015,Hossen2016a,Hossen2016b,Hossen2017,Sahu2012b}
have considered non-extensive ions, and have also used three possible ranges of $q$
(i.e., $q=$ negative, $q=$ positive but less than $1$, and $q=$ positive but grater than 1)
for numerical study of the nonlinear electrostatic structures associated with DAWs in an EDDP,
and have also identified that the possible existence of such kind of non-extensive ions along
with dust grains in cometary tail \cite{Zaghbeer2014,Hossen2016a,Hossen2016b},
upper mesosphere \cite{Jahan2019,Zaghbeer2014}, magnetosphere of Jupiter \cite{Jahan2019},
and Saturn's F-ring \cite{Bains2013,Sahu2012b}, etc.

\begin{figure}[t!]
\centering
\includegraphics[width=80mm]{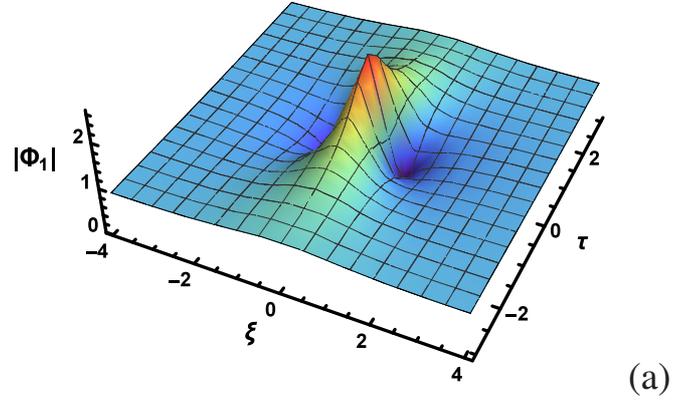}
 \Large{(a)}
\vspace{0.5cm}
\includegraphics[width=80mm]{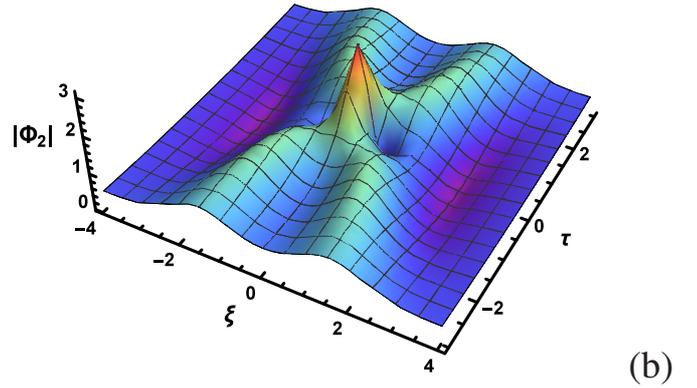}
\Large{(b)}
\caption{Profile of the (a) first-order rational solution; (b) second-order rational solution when other parameters are
$e_1=0.07$, $e_2=0.007$, $e_3=0.6$, $e_4=2.0$, $q=2$, $k=0.5$, and $\omega_f$.}
 \label{1Fig:F11}
\end{figure}
\begin{figure}[t!]
\centering
\includegraphics[width=80mm]{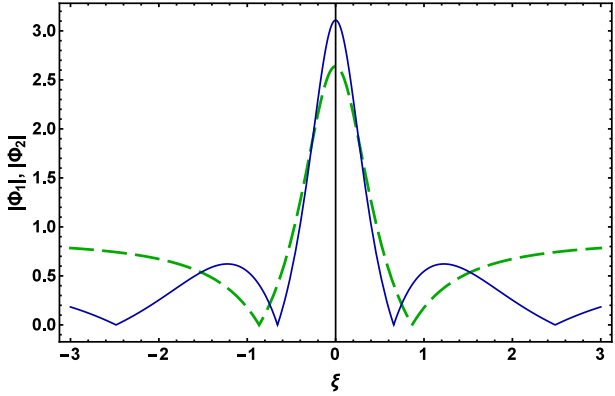}
\caption{Profile of the first-order (dashed green curve) and second-order (solid blue curve) rational solutions when other parameters are $e_1=0.07$, $e_2=0.007$, $e_3=0.6$, $e_4=2.0$, $\tau=0$, $q=2$, $k=0.5$, and $\omega_f$.}
\label{1Fig:F12}
\end{figure}
The space and time evolution of the first-order and second-order rational solution of the
NLSE \eqref{1eq:42} can be observed from Figs. \ref{1Fig:F11}(a) and \ref{1Fig:F11}(b), respectively.
Figure \ref{1Fig:F12} indicates the first-order and second-order solution at $\tau=0$, and it is
clear form this figures that (a) second-order rational solution has double structure compared with
first-order rational solution; (b) the height of the second-order rational solution is always greater
than the first-order rational solution; (c) the potential profile of the second-order rational solution
becomes spiky (i.e., the taller height and narrower width) than the  first-order rational solution;
(d) the second (first) order rational solution has four (two) zeros symmetrically located on the
$\xi$-axis; (e) the second (first) order rational solution has three (one) local maxima.

The existence of highly energetic RWs has already been confirmed
experimentally \cite{Chabchoub2011,Chabchoub2012,Bailung2011,Kibler2010} and theoretically \cite{Shalini2015}.
The second-order solution was experimentally observed by Chabchoub \textit{et al.} \cite{Chabchoub2012} in a ``Water Wave Tank''
and the experimental result regarding the amplification is a nice agreement with the theoretical result.
Bailung \textit{et al.} \cite{Bailung2011} demonstrated an experiment in a multi-component plasma to observe RWs, and
found a slowly amplitude modulated perturbation undergoes self modulation and gives rise to a high amplitude localized pulse.
Rogue waves also observed in fiber optics \cite{Kibler2010}.
\section{Conclusion}
\label{1sec:Conclusion}
In our present article, we have investigated the characteristics of the amplitude
modulation of DAWs by using a NLSE, which are successfully
derived by employing the standard RPM in a EDDP composed of non-extensive ions, negatively
and positively charged warm dust grains. In the formation and propagation of DAWs, the moment of inertia is provided
by the mass of the heavier components (adiabatic warm positive
and negative dust grains) and restoring force is provided by the thermal pressure of the lighter component
(non-extensive ion) of the plasma medium.
So, each of the plasma components of the plasma medium
provides a great contribution to the formation and propagation
of the DAWs in three components EDDP medium. The observational data have disclosed the ubiquitous existence
of non-extensive ions as well as positive and negative dust grains
 in astrophysical environments such as cometary tail \cite{Zaghbeer2014,Hossen2016a,Hossen2016b},
Saturn's F-ring \cite{Bains2013,Sahu2012b}, upper mesosphere \cite{Jahan2019,Zaghbeer2014}, and
magnetosphere of Jupiter \cite{Jahan2019} as well as laboratory situations such as laser-matter
interaction \cite{Shahmansouri2013}. Hence, the implications of our results should be useful to understand
the mechanism of MI of the DAWs and the formation of DARWs as well as envelope solitons in the modulationally unstable parametric regime, which is
determined by the sign of P and Q of the standard NLSE, in a three components EDDP medium such as
cometary tail \cite{Zaghbeer2014,Hossen2016a,Hossen2016b}, Saturn's F-ring \cite{Bains2013,Sahu2012b},
upper mesosphere \cite{Jahan2019,Zaghbeer2014}, and magnetosphere of Jupiter \cite{Jahan2019}
as well as laboratory situations such as laser-matter interaction \cite{Shahmansouri2013}.
It may be noted here that the gravitational and magnetic field effects are very important
but beyond the scope of our present work. In future and for better understanding, someone can study the nonlinear
propagation in a three components EDDP medium by considering the gravitational and magnetic field effects.

\end{document}